# Evidence for Polariton-Mediated Biexciton Transition in Single-Walled Carbon Nanotubes

*Jan M. Lüttgens[1], Zhuoran Kuang[1,2], Nicolas F. Zorn[1], Tiago Buckup\*[1,2], and Jana Zaumseil\*[1]*

[1]Institute for Physical Chemistry, Universität Heidelberg, D-69120 Heidelberg, Germany

[2]Centre for Advanced Materials, Universität Heidelberg, D-69120 Heidelberg, Germany

**Corresponding Authors**

*E-mail: zaumseil@uni-heidelberg.de

*E-mail: tiago.buckup@pci.uni-heidelberg.de

ABSTRACT

Strong coupling of excitonic resonances with a cavity gives rise to exciton-polaritons which possess a modified energy landscape compared to the uncoupled emitter. However, due to the femtosecond lifetime of the so-called bright polariton states and transient changes of the cavity reflectivity under excitation, it is challenging to directly measure the polariton excited state dynamics. Here, near-infrared pump-probe spectroscopy is used to investigate the ultrafast dynamics of exciton-polaritons based on strongly-coupled (6,5) single-walled carbon nanotubes in metal-clad microcavities. We present a protocol for fitting the reflectivity-associated response of the cavity using genetic algorithm-assisted transfer matrix simulations. With this approach are able to identify an absorptive exciton-polariton feature in the transient transmission data. This feature appears instantaneously under resonant excitation of the upper polariton but is delayed for off-resonant excitation. The observed transition energy and detuning dependence point toward a direct upper polariton to biexciton transition. Our results provide direct evidence for exciton-polariton intrinsic transitions beyond the bright polariton lifetime in strongly-coupled microcavities.

Exciton-polaritons are hybrid light-matter quasiparticles that are formed when confined light interacts strongly with an excitonic resonance, *e.g.*, in a metal-clad microcavity.[1, 2] In this strong-coupling regime, mixed photon-exciton eigenstates arise that are called upper and lower polariton (UP and LP, respectively) with regard to the energy of the underlying excitonic transition. For molecular emitters, strong-coupling offers new means to manipulate photophysical properties without chemical modification. Polariton states are readily tunable due to their photonic component and thus may provide alternative pathways for photoexcited processes, *e.g.*, photocurrent-generation[3-5] and spin conversion,[6-10] which has led to the efflorescence of the field of molecular polaritonics.[11] The delocalized photonic character of polaritons was also shown to improve energy transfer between spatially separated molecules[12, 13] and long-range energy transport[14, 15] in disordered molecular systems. To further improve and expand these applications of molecular polaritonics, it is essential to understand the fundamental properties of polaritonic states, especially their excited state dynamics.

The steady state properties of strongly-coupled macroscopic molecular systems are well-understood and can be modelled as coupled oscillators.[16] In such bulk systems, the Rabi splitting ($\hbar\Omega$) is a measure for the coupling strength, which scales with the square root of the number ($N$) of emitters (molecules) ($\hbar\Omega \propto \sqrt{N}$) in the cavity. One emitter state and the photon mode form the bright UP and LP modes, while ($N$ - 1) emitters form degenerate dark polariton states, that are essentially molecular in nature.[17] The polariton steady state and the observable time-dependent fluorescence can be explained by population of the bright polaritons from the so-called exciton reservoir by scattering[18] or non-adiabatic coupling (radiative pumping).[19-22] The exciton reservoir consists of dark polariton states and uncoupled molecular states.[23]

Despite of these general insights, the specific conversion dynamics between exciton-polariton states (UP and LP) and molecular states have remained elusive.[24] This shortcoming is mainly due to a lack of suitable ultrafast spectroscopy methods. Especially in metal-clad microcavities the intrinsic polariton lifetime, as derived from linewidth, ranges from a few to tens of femtoseconds.[25] This timescale is comparable to the shortest available laser pulses in the visible range and therefore not directly resolvable with conventional pump-probe measurements. Yet, the majority of the few studies on transient transmission and reflectivity of strongly-coupled microcavities show that the overall response of the polariton modes is long-lived and emitter-like.[26-29] This effect can be attributed to the pump induced bleach of the underlying emitter, which leads to a transient change of the emitter's complex refractive index and an associated transient transmission and reflectivity response, that directly follows the emitter dynamics.[27, 28] Depending on excitation conditions, thermal reflectivity of the metal mirrors[30] or transient changes of the cavity materials, such as the refractive index or layer thickness,[31] may further contribute to the cavity transient transmission and reflectivity response. Owing to these dominant, non-polaritonic cavity transient effects, the evolution of the polariton population has not yet been extracted from pump-probe experiments with any certainty. Multiscale molecular dynamics simulations have shown that the short-lived bright polariton states may transfer population to the long-lived dark polariton states reversibly, thus elongating the bright polariton lifetime.[32] A two-dimensional Fourier transform spectroscopy study could confirm coherent energy exchange between UP, LP and exciton reservoir during the first 100 to 150 fs after excitation.[24] However, for dynamics that take place after dephasing between photon and molecular components, no polariton intrinsic features, *i.e.*, spectral signatures that are not merely the result of the transient change of the emitter's refractive index, have yet been observed.[26-29]

Here, we investigate the ultrafast dynamics of polariton states by fitting the emitter-bleach associated transient response of the system to identify absorptive features, that are directly connected to the evolution of the polariton population. We study these effects for metal-clad microcavities with polymer-sorted (6,5) single-walled carbon nanotubes (SWCNTs). (6,5) SWCNTs are one-dimensional semiconductors with a bandgap of about 1.27 eV, absorbing and emitting near-infrared light.[33] Purified semiconducting carbon nanotubes have emerged recently as an excellent material to create exciton-polaritons in planar microcavities[21, 34-36] as well as in plasmonic lattices[37, 38] due to their large oscillator strength, stable excitons with large binding energies, narrow absorption ($E_{11}$, $E_{22}$ *etc.*) and emission (only $E_{11}$) peaks with a small Stokes shift. Even ultra-strong coupling, electrical tuning and excitation of these SWCNT polaritons is easily achieved.[35, 39, 40]

Furthermore, SWCNTs exhibit intriguing photophysics and dynamics that have been studied in detail by transient absorption and other time-resolved spectroscopies.[41-44] The large Coulomb interactions in one-dimensional nanotubes and the reduced dielectric screening enable many-body bound states such as trions or biexcitons. Trions are charged excitons that occur in doped nanotubes[45-47] but can also be created optically at high excitation intensities.[48] For (6,5) SWCNTs they show red-shifted (by 170-190 meV from the $E_{11}$) absorption and emission. Biexcitons are bound states of two excitons and are often observed in zero-dimensional systems (*e.g.*, quantum dots[49]) as well as nanotubes.[50, 51] They are usually produced by the collision of two excitons, but can also be created from an existing exciton population by absorption of additional photons. Hence, they are observable as an induced absorption feature in transient absorption measurements, which is red-shifted from the $E_{11}$ exciton by about 130 meV for (6,5) SWNTs as shown previously.[52, 53]

By comparing the ultrafast transient transmission spectra of strongly-coupled (6,5) SWCNTs in metal-clad cavities with genetic-algorithm assisted transfer matrix simulations, we identify an exciton-polariton absorptive feature. Under resonant UP excitation we observe this feature instantaneously at an energy matching a direct UP to biexciton transition. For non-resonant excitation, the proposed UP to biexciton transition is retained but occurs at later times. This indicates efficient population transfer between UP and dark polariton states for suitable detunings with strong overlap between UP and dark states.

RESULTS AND DISCUSSION

**Cavity-properties and steady-state data.** Monochiral, polymer-sorted (6,5) SWCNTs were used as the active excitonic material in a metal-clad Fabry-Pérot cavity. These (6,5) nanotubes exhibit distinct absorption lines corresponding to the transitions to the excitonic states $E_{11}$ (1.232 eV) and $E_{22}$ (2.160 eV) with emission only from the $E_{11}$ state (see reference film absorbance and emission in **Figure S1, Supporting Information**). For the fabrication of the cavity (schematically shown in **Figure 1a**), a dense and homogeneous (6,5) SWCNT film (thickness 30-40 nm) was spin-coated from a highly concentrated dispersion onto a gold-coated (30 nm) glass substrate with an aluminium oxide ($AlO_x$) spacer (120 nm, see Methods for details). After formation of the nanotube layer another $AlO_x$ spacer (120 nm) and a top gold (30 nm) mirror were deposited. A (6,5) SWCNT reference film was deposited under the same conditions but without the mirrors and protected with (120 nm) $AlO_x$ for comparability. The cavity tuning was determined by the spacer layer thickness. By concentrating the nanotubes at the field maximum of the $\lambda/2$ cavity the number of uncoupled emitters was reduced.

**Figure 1b** depicts the *p*-polarized angular dispersion of reflectivity (R) and photoluminescence (PL) for such a microcavity as recorded by Fourier imaging (see Methods for details). The reflectivity data exhibits a clear anti-crossing at the coupled (6,5) SWCNT exciton ($E_{11}$, 1.232 eV) indicative of strong coupling. The UP and LP modes were fitted to the reflectivity data with the coupled oscillator model (see Methods for details), yielding a Rabi splitting of 80 meV. The quality-factor estimated from the LP linewidth was Q ≈ 25, corresponding to an ultrashort cavity lifetime of 15 fs. For the detuning chosen here (-140 meV), the UP strongly overlaps with the $E_{11}$ absorption leading to efficient coupling to the polariton dark states (DS),[32] whereas the LP mainly overlaps with the (6,5) SWCNT emission sidebands (**Figure S1b**, **Supporting Information**). As a consequence, no emission is observed from the UP,[54] while the LP emits efficiently (see PL in **Figure 1b**) for energies with maximum overlap between LP and the (6,5) SWCNT photoluminescence tail and sidebands, as reported previously.[21] Note that the PL decay from the LP follows approximately the PL decay dynamics of (6,5) SWCNTs without a cavity (see **Figure S2, Supporting Information**), because the LP fluorescence decay is limited by the slow population from the dark states, that possess a similar lifetime as the weakly coupled SWCNT excitons.

**Figure 1c** shows the steady state transmission (*T*) at normal incidence of the same cavity, given as a reference to identify the transient features of the UP and LP mode in the following. Note that the LP transmission peak (**Figure 1c**) appears broader than the LP in reflectivity due to the color scale in **Figure 1b**. A superposition of transmission and reflectivity data can be found in the **Supporting Information** (**Figure S3**).

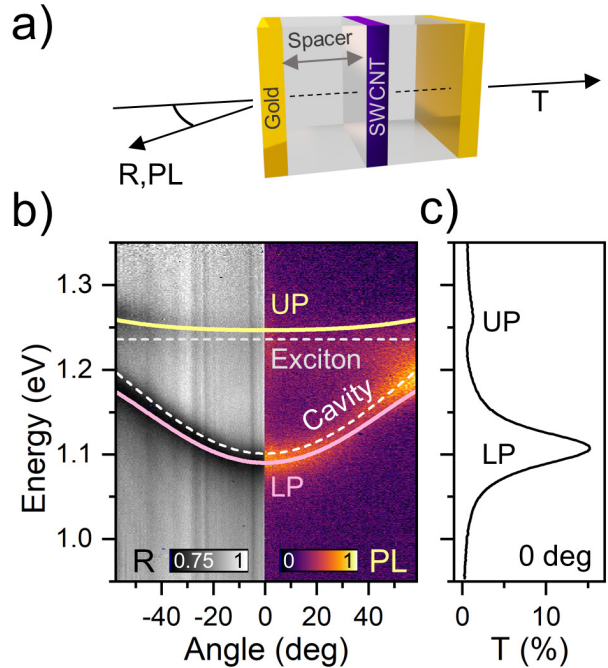

**Figure 1.** Strongly coupled (6,5) SWCNTs in a metal-clad microcavity. (a) Schematic of the sample structure and detection geometry. (b) Angle-resolved steady state reflectivity (R) and photoluminescence (PL) of (6,5) SWCNTs in a planar cavity. The polariton modes (UP and LP) and cavity dispersion of a coupled oscillator fit to the reflectivity data are indicated as colored lines. (c) Steady state transmission (*T*) of the same cavity at normal incidence.

**Transient transmission data.** The polariton dynamics were investigated by applying pump-probe spectroscopy in a transmission geometry. The transient absorption (TA) data of a (6,5) SWCNT film sandwiched between two aluminium oxide layers (**Figure S4**, **Supporting Information**) serves as a reference for the emitter dynamics in the weak coupling regime. Cavity-embedded SWCNTs and reference SWCNTs were subjected to the same excitation power. For this, the reference was excited with lower pulse energies (2 nJ) and the cavity with higher pulse energies (30 nJ) to account for excitation losses (~ 93 %) at the top-mirror of the cavity sample. Before analyzing the experimental transient transmission (TT) data, it is important to consider the variations in reflectivity during and after excitation, as the corresponding change in sample transmission is convoluted with the change in absorption.

The transient change in reflectivity of the reference sample is negligible as the PFO-BPy matrix and AlO$_x$ layers are transparent at the excitation wavelength (575 nm) and the excitation density (4 µJ cm$^{-2}$) is too low for non-linear processes such as multi-photon absorption, which could affect the refractive index and consequently the reflectivity. For the strongly-coupled cavity, the reflectivity contributions have to be separated from polariton excited state absorption (ESA). For metal-clad cavities the dominant contribution is thermal excitation of the cavity mirrors,[30] especially the top mirror, which interacts with the unattenuated pump-pulse. For the applied experimental parameters this effect is small in our samples (< 1 mOD, see **Figure S5**, **Supporting Information** together with a discussion of potential thermal expansion of the adjacent spacer layer). Furthermore, about 93% of the pump pulse is reflected at the top mirror, the pump-fluence incident on the dielectrics is again low (4-40 µJ cm$^{-2}$) and non-linear effects can be disregarded. Note that under different excitation conditions, the dielectric layers might be affected by the excitation pulse.[31]

Based on the reasoning above, the change in reflectivity is dominated by the bleach of the SWCNTs, leading to an emitter-like component in the transient response of LP and UP. This emitter-like component of the corresponding cavity transient transmission was determined by transfer-matrix (TM) simulation. The transient change in the complex refractive index of the emitter layer was extracted from the transient absorption of the SWCNT reference using Kramers-Kronig relation[27] and employed to calculate the optical response of the strongly-coupled sample for each pump-probe delay (for details see **Supporting Information**, **Figure S6**). To account for small variations in the SWCNT film and in the cavity layers, a genetic algorithm was used to optimize the structural input parameters for the TM simulation (*e.g.*, layer thicknesses and layer roughness) to fit the investigated area of the cavity sample (for details see **Supporting Information, Figures S6 – S7, Tables S1 and S2**). By comparing experimental and simulated TT data, we were able to identify intrinsic polariton features, as they are not contained in the TM simulation, which was purely based on the transient refractive index change of the weakly coupled emitter.

**Figures 2a** and **2b** depict the experimental TT data, given as $\Delta T = -\lg(T/T_0)$, of the strongly-coupled cavity described above when excited off-resonantly at the SWCNT $E_{22}$ exciton transition. The spectrum exhibits two positive and one negative component that decay at similar rates over a few hundred picoseconds. The region from 1.05 to 1.18 eV can be attributed to the LP ground state response and the region from 1.18 to 1.3 eV to the UP ground state response. These features can be interpreted as follows. The pump-pulse promotes ground-state population to the $E_{22}$ exciton manifold and the resulting ground state bleach reduces the absorption at the $E_{11}$ transition leading to a reduction in Rabi splitting as shown in **Figure S8**, **Supporting Information**. For the UP mode, this leads to an increase in transmission (purple) below 1.25 eV and a decrease in transmission (orange) above 1.25 eV as the UP mode is shifted towards lower energies. The same process occurs for the LP around 1.13 eV with inverse signs. After arrival of the pump-pulse, the ground state population recovers and the Rabi splitting increases again, which is monitored by the probe-pulse. This process produces the characteristic derivative-like lineshape in the transient polariton response of the polariton ground state (compare **Figures 2b, 2d and 2f** with **Figure S8**, **Supporting Information**), which is ubiquitous in strongly-coupled Fabry-Pérot cavities.[27, 28, 55] Note that within the probed time-window the LP shifts exclusively to lower energies and the UP to higher energies, approaching their equilibrium positions observed in steady-state transmission.

The TM simulation (see **Figures 2e** and **2f**) captures this behavior quite accurately without invoking polariton excited state dynamics.

Further comparing the experimental data (**Figure 2b**) to the simulation (**Figure 2f**), we find slightly broader lineshapes for the former, mostly because in the simulation the interface roughness was modelled by a global scalar scattering parameter and scattering within the layers was neglected. Despite the overall good agreement between the TM scheme and experiment at later times (> 1 ps), larger discrepancies are found at early times (< 1 ps, see **Figures S9a** and **S9b**, **Supporting Information**). One cause for this mismatch may be the more prominent role of equilibration between UP, LP and dark polariton states, which has been predicted to evolve over the first few hundred femtoseconds (100-200 fs) for direct excitation of the emitter in the case of small molecules.[32] Another cause for this mismatch is the omission of coherent interaction between pump- and probe-pulse at early times (<100 fs). At later times we expect the mismatches to arise from the intrinsic polariton dynamics, not included in the simulation.

Importantly, there is an unexpected red-shifted shoulder of the LP response in the experimental data of the strongly-coupled cavity, which is completely absent in the simulation (compare **Figure 2a** with **2e** and **Figures S9c** and **S9d**, **Supporting Information**). The origin of this shoulder cannot be explained by a transient change of transmission. Firstly, the LP shifts exclusively to lower energies towards its equilibrium position. Hence, transitions lower in energy than the steady-state LP transition indicate ESA. Secondly, weakly coupled SWCNT-related ESA, that may be observable through the red flank of the LP is accounted for by the simulated TT spectrum. This is shown in **Figure S9c** (**Supporting Information**), where the experimental TT spectrum is superimposed on to the simulation and steady-state transmission. Hence, the observed shoulder must originate from polariton ESA.

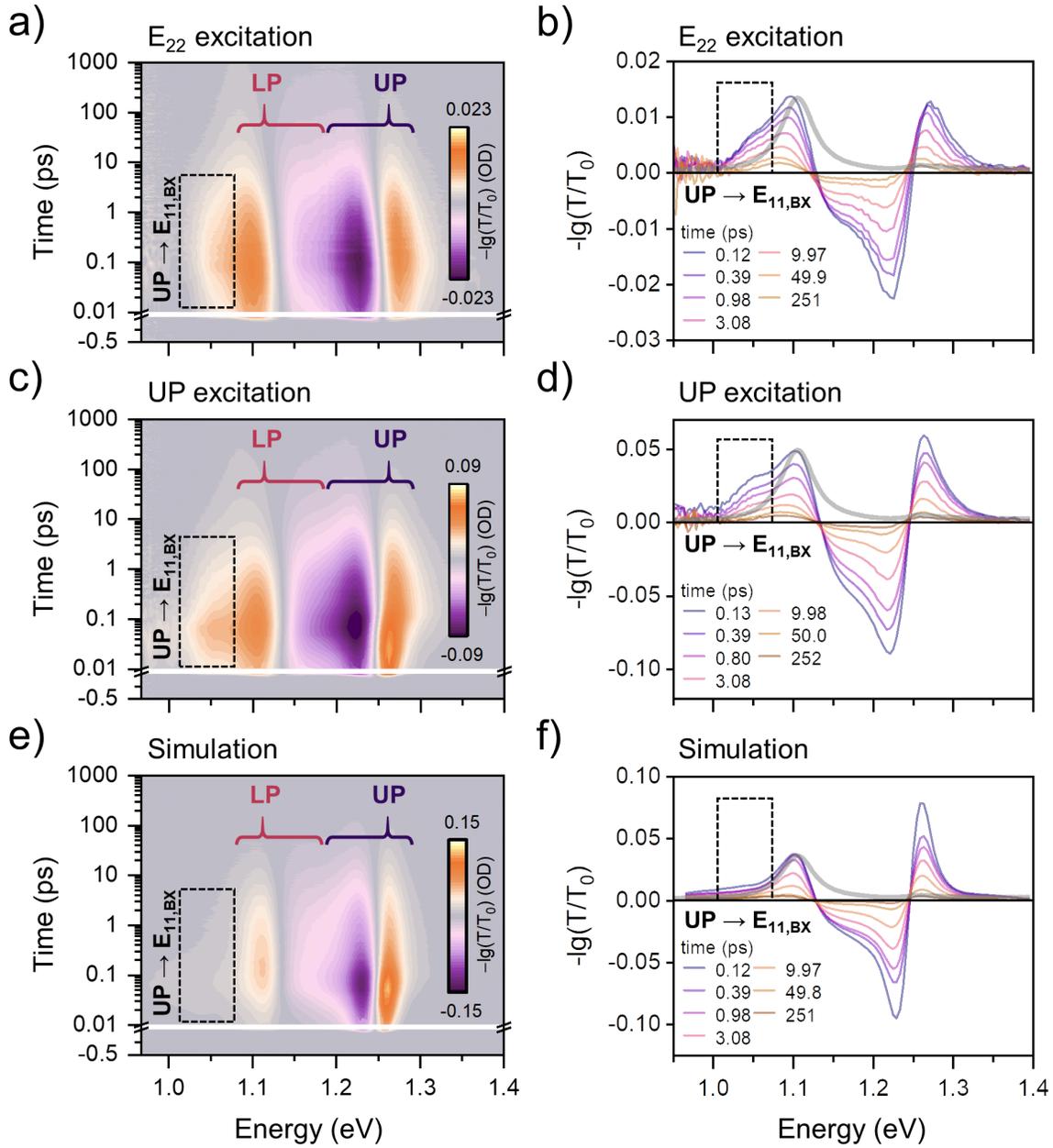

**Figure 2.** UP to biexciton ($E_{11,BX}$) transition as an intrinsic polariton ESA feature. (a,b) Transient transmission (TT) data of a strongly-coupled SWCNT microcavity excited off-resonantly at the $E_{22}$ exciton. (c, d) TT data of the same cavity resonantly excited at the UP mode. (e,f) Simulated TT data based on the transient absorption of a SWCNT reference film. The spectral regions of LP (red) and UP (dark purple) responses are indicated with the bracket tip pointing at the respective steady state peak positions. The proposed UP to biexciton transition (UP → $E_{11,BX}$) is indicated by a dashed outline. The grey solid lines represent the steady state transmissions of the cavities (b, d, f).

The spectral position of the shoulder (1.05 eV) equals approximately the energy separation between the UP and the $E_{11}$ biexciton (BX) of the (6,5) SWCNTs, which we estimated to be around 1.07 eV ($2E_{11}(1.23\ eV) - E_{UP}(1.26\ eV) - E_{Bind}^{BX}(0.13\ eV)$) for this cavity sample. The biexciton binding energy ($E_{Bind}^{BX}$) was determined by Yuma *et al.* for surfactant dispersed (6,5) SWCNTs in water.[53] Note that for PFO-BPy wrapped SWCNTs, the $E_{11}$ energy is shifted to lower energies by 30 meV, as is the biexciton absorption ($E_{11}(1.23\ eV) - E_{Bind}^{BX}(0.13\ eV) = 1.1\ eV$, see region around 1.1 eV for reference film, **Figure S4**, **Supporting Information**). We can exclude a transition into the SWCNT trion, because trions are only observed in the TA of chemically doped PFO-BPy wrapped SWCNTs[47] and were also absent in the TA of the reference film. Consequently, the most dominant excited state transition in our system should be the $E_{11}$ to biexciton transition and we therefore interpret the observed red-shifted shoulder in the cavity TT spectrum as a direct UP to biexciton transition.

To further test this assignment, the cavity sample was excited resonantly at the UP energy. **Figures 2c** and **2d** depict the corresponding TT data. Apart from the expected UP and LP ground state features, the same red-shifted shoulder appears slightly more pronounced and at earlier times. In contrast to off-resonant excitation (**Figures 2a** and **2b**), the UP is directly populated by the pump-pulse, which is consistent with an immediate onset of the proposed biexciton absorption.

The absorptive feature is also observed for direct excitation of the LP and with a slight delay of 45 fs (see **Figures S10a** and **S10b**, **Supporting Information**). To the best of our knowledge, population transfer from the LP to the UP has not been reported for pure LP excitation (note that the pump width at the LP energy is about 25 meV). Although phonon-assisted nonadiabatic transitions among polariton states have recently been reported,[56] the question of how the large energy gap of -140 meV between LP and DS could be surpassed remains elusive and will be subject of future work.

As seen in **Figures 2a** and **2b**, the red shoulder indicates that there is UP population even after 1 ps, far beyond the intrinsic UP lifetime (~ 15 fs). Hence, the UP population must be replenished from a long-lived state, which should be the exciton reservoir or dark-states. The interpretation of the UP population being the rate limiting step for the observed UP to biexciton transition is similar to the explanation for the experimentally observed long-lived LP photoluminescence decays.[19-21] We tested this hypothesis by measuring a cavity with a thicker oxide spacer, *i.e.*, a larger detuning, and a cavity with a thinner oxide spacer, *i.e.*, a smaller detuning. For the larger detuning (-184 meV) we could still observe a red-shifted shoulder for UP excitation (see **Figure S11a, Supporting Information**), as the overlap between UP and exciton reservoir was not changed significantly compared to the data shown in **Figure 2**. However, for a smaller detuning (-33 meV), *i.e.*, reduced overlap between the UP and the exciton reservoir, the shoulder vanished (see **Figure S11b, Supporting Information**).

**Decay associated difference spectra (DADS).** The experimental and simulated pump-probe data shown in **Figure 2** can be further analyzed by global analysis. Without assuming any detailed kinetic model, the data is fitted globally (*i.e.*, all wavelengths are fitted simultaneously) with a sum of exponentials where the amplitudes depend on the wavelength. The number of exponentials is usually equal to the number of spectrally different components, *e.g.*, electronic states, molecules, *etc*. The amplitude of each exponential is called decay associated difference spectrum (DADS).[57] All three datasets in **Figure 2** could be described using 5 decays (1-5) and an offset accounting for measurement noise.

**Figure 3** shows the respective DADS with the corresponding time constants ($k_1 - k_5$) being summarized in **Table 1**. The global analysis again reveals the UP to biexciton transition (UP → $E_{11,BX}$) peak at around 1.05 eV for excitation at $E_{22}$ or UP. This is especially clear in the second decay component $k_2$. In case of UP excitation, the first two decays are about twice as fast as for $E_{22}$ excitation. We attribute the slower decays for $E_{22}$ excitation to the delay caused by the required internal relaxation to the $E_{11}$ state (~ 100 fs)[47] and subsequent population of the UP by the dark polariton states. This evolution of the proposed UP to biexciton transition approximately follows the evolution of the LP feature (at 1.1 eV in **Figure 3a** and **3b**). As mentioned earlier, the LP feature is a direct consequence of the SWCNT ground state bleach. Hence, the simultaneous spectral evolution of the UP and the LP feature indicates that the polariton-mediated biexciton transition follows the decay of the ground state bleach. This is in agreement with previous experiments on weakly coupled (6,5) SWCNTs, in which the biexciton population followed the exciton population[53] and further corroborates the biexciton character of the observed transition at 1.05 eV. Note that the $k_2$ and $k_3$ decay rates still contain the ground state response. The UP to BX feature visible in the $k_2$ and $k_3$ spectra shows, that for these time constants it is a contribution, not that the UP to BX feature decays with precisely these rates.

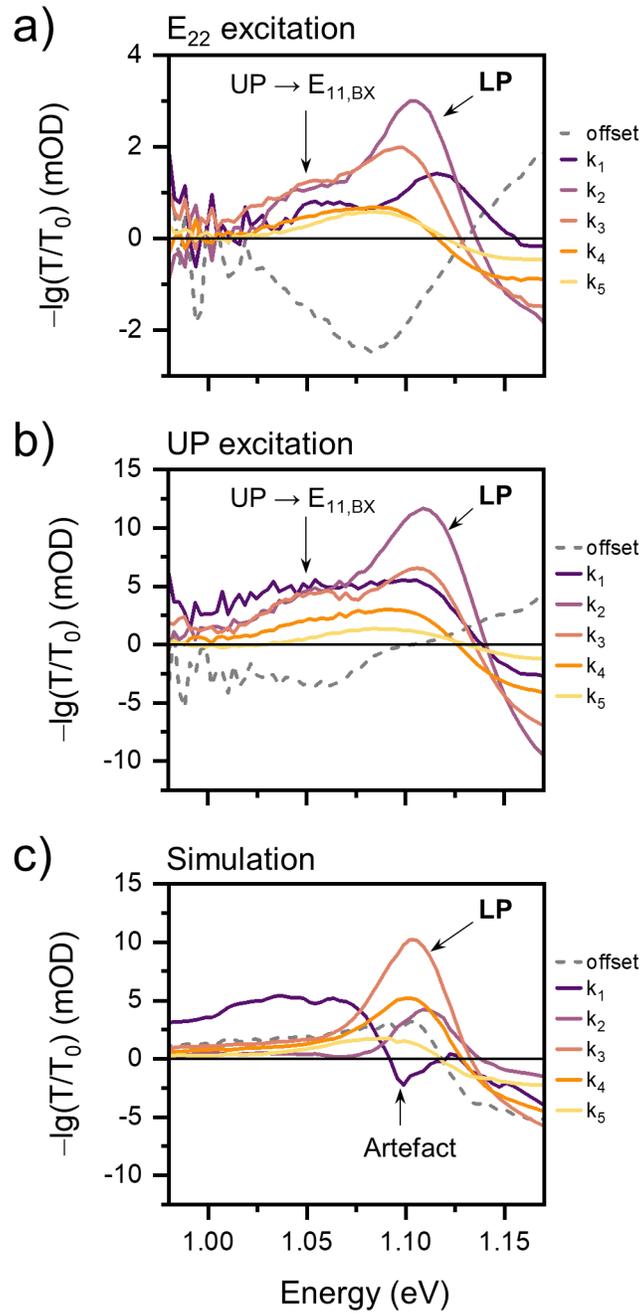

**Figure 3.** Biexciton absorption feature in decay associated difference spectra (DADS). (a) Strongly-coupled (6,5) SWCNT microcavity excited off-resonantly at $E_{22}$ (a), excited resonantly at the UP mode (b) and for a corresponding TM simulation (c). The extracted time constants ($k_1$ – $k_5$) and lifetimes can be found in **Table 1**.

**Table 1.** Summary of lifetimes (1/k) obtained by the global fit.

| Lifetime component | $1/k_1$ (ps) | $1/k_2$ (ps) | $1/k_3$ (ps) | $1/k_4$ (ps) | $1/k_5$ (ps) |
|---|---|---|---|---|---|
| $E_{22}$ excitation | $0.53 \pm 0.06$ | $2.0 \pm 0.2$ | $10 \pm 1$ | $47 \pm 6$ | $380 \pm 40$ |
| UP excitation | $0.20 \pm 0.02$ | $1.15 \pm 0.05$ | $5.3 \pm 0.2$ | $38 \pm 2$ | $543 \pm 40$ |
| Simulation | $0.127 \pm 0.003$ | $0.78 \pm 0.03$ | $4.06 \pm 0.06$ | $20.9 \pm 0.2$ | $328 \pm 3$ |

The polariton intrinsic nature of the absorptive feature at 1.05 eV (**Figure 2a-d**) as an UP to biexciton transition is further corroborated by the global analysis of the simulated data (see **Figure 3c**). As described above, the simulated data does not contain any polariton intrinsic features and therefore, the DADS shown in **Figure 3c** lack any feature at 1.05 eV. It is important to note that the TM simulation does not include any coherent interaction at early delay times between pump and probe ($\Delta t < 0.5$ ps) and the interpretation of the first component $k_1$ of the simulated data must be considered carefully. For example, the negative feature at the LP position (~ 1.1 eV, **Figure 3c**) is an obvious indication for the $k_1$ component to be unphysical. The slower components ($k_2$ - $k_5$, **Figure 3c**), however, are not affected by artefacts, and can be interpreted safely. Note that this only applies to the simulated data. The absence of the absorptive transition feature in these components of the simulated data corroborates that it is not caused by a change in transmission of the cavity stack and thus, is indeed intrinsic to the polariton dynamics.

For the lifetimes calculated from the $k_2$ to $k_5$ components of the simulated data, we find similar values as for UP excitation (see **Table 1**). As the dynamics of the simulated data arise directly from the dynamics of the weakly-coupled SWCNTs, the timescales covered by $k_2$ to $k_5$ are dominated by the emitter dynamics with regard to the polariton ground state response (compare

LP feature in **Figure 3**) in agreement with previous reports.[27, 28] The consistently slower decays of $k_3$ to $k_4$ for $E_{22}$ excitation are likely not polariton-related and an intrinsic feature of SWCNTs. For polymer-wrapped (6,5) SWCNTs in tetrahydrofuran it was observed, that the overall decay times were slightly elongated for excitation to higher excitonic levels such as the $E_{22}$ or $E_{33}$.[47] The $k_5$ values should be essentially similar, however, the increased noise level at later times makes the fit less reliable and the stated fit error likely underestimates the real uncertainty.

**Kinetic model and implications for UP population.** We now assess the population of the UP state and the efficiency of the proposed polariton-mediated UP to biexciton transition. **Figure 4a** presents the kinetic model based on the previous discussion including the ground state (GS), different excited ($E_{11}$, $E_{22}$, BX) and polaritonic (UP, LP, DS) states of strongly-coupled (6,5) SWCNTs. **Figure 4b** shows the respective normalized time traces at the transition energy of the absorptive feature for resonant (at UP) and off-resonant (at $E_{22}$) excitation, and the exciton to biexciton transition of the reference film. For resonant excitation at the UP, the maximum population is reached after around 60 fs within the experimental time-resolution (instrument response function ~90 fs). This indicates, that the transition is connected to the UP population, of which a significant fraction should decay within the instrument response time. For off-resonant excitation at the $E_{22}$ transition, the maximum population is reached after around 130 fs, indicating, that the transition is delayed by internal relaxation and population of the DS and subsequently the UP. The fast decay of the biexciton population within the first 150 fs observed for UP excitation, is absent for off-resonant $E_{22}$ excitation. In the former scenario, population is injected directly into the UP, from which the product states (which we propose to be the biexciton) can be populated efficiently at early times. After 150 fs, considerable UP population is lost due to relaxation into the dark states as inferred from PL measurements (see **Figure 1b**). From then on, the proposed UP to biexciton transition feature evolves similar to off-resonant $E_{22}$ excitation. Thus, we assume, that the UP to dark state relaxation must be dominant, while the UP is constantly repopulated from the dark states at a slower rate. The resulting finite population can explain the slow decay of the UP to biexciton transition, at rates similar to the biexciton feature in the reference film.

As mentioned earlier, the connection between UP and dark states can be tested, by changing the overlap between the two states, *i.e.*, the detuning of the cavity. For more negative detunings (still with similar overlap between UP and dark states) the UP to biexciton transition is still observed, whereas for less negative detuning the transition vanishes (**Figure S11, Supporting Information**). The latter also underlines, that the DS to UP transition is entropically and enthalpically disfavored, if the overlap between DS and UP is small and the energy barrier is high.[58] Yet for a highly negative detuning, the DS to UP transition becomes competitive compared to the reverse process.

Due to the spectral overlap between UP to BX transition and LP ground state response, the kinetics of the UP to BX cannot be extracted from the experimental cavity data of **Figure 4**. However, for a DS to UP rate of $(150 \text{ fs})^{-1}$ and a bright polariton lifetime of $(15 \text{ fs})^{-1}$ the kinetic model proposed above yields an evolution of the UP population, that is similar to the experimental time traces of the UP to BX transition (see **Figure S12, Supporting information**). The kinetic model also shows, that the UP population may have a lifetime in the picosecond range, if efficient population exchange with the DS is possible.

The efficiency of the proposed UP to biexciton transition can be estimated by comparing the absolute signal intensities between cavity and reference sample at the respective biexciton features (see **Supporting Information, Figure S13**). This comparison reveals that the UP to biexciton transition feature is three times more intense for the strongly-coupled sample under $E_{22}$ excitation than for the reference. For resonant excitation of the UP, the transition efficiency can be increased further up to four-fold compared to off-resonant excitation as the dark states are bypassed. Note that this is the case even though $E_{22}$ excitation should be 17% more efficient based on the calculated electromagnetic field intensity inside the cavity (see **Supporting Information, Figure S14**).

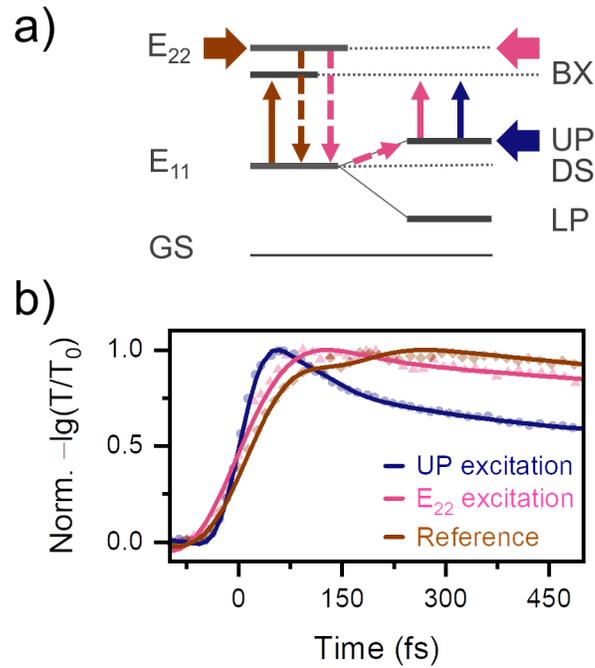

**Figure 4.** (a) Kinetic model for polariton assisted biexciton transition by resonant (UP, blue) and off-resonant ($E_{22}$, pink) excitation in comparison to the SWCNT reference (brown). (b) Fitted time-traces of the biexciton absorption in the reference film (extracted at exciton to biexciton energy) and strongly coupled cavity (extracted at the transition energy of the absorptive feature) with color-coding for the respective experimental conditions.

## CONCLUSION

By analyzing and comparing the transient transmission response of strongly-coupled (6,5) SWCNTs in metal-clad microcavities with genetic algorithm-assisted transfer matrix simulations, we were able to identify an intrinsic exciton-polariton absorptive feature. The transition energy and kinetics of this feature point towards a direct upper polariton to biexciton transition. This transition is three times more efficient than the biexciton transition in a weakly-coupled reference. It is present for both resonant UP and non-resonant excitation at $E_{22}$, however, only for detunings at which the UP can efficiently exchange population with the dark states. The kinetics of the observed polariton-mediated absorptive feature imply, that the UP can be repopulated by the dark states to a significant extent, thus increasing the UP population lifetime to the picosecond range despite the ultrashort intrinsic UP lifetime. These observations underline the need for tracing the polariton population in such systems directly. Transfer matrix simulations assisted by global optimization methods, as shown here for a genetic algorithm, can be powerful tools to analyze the corresponding data with greater reliability and thus help to identify and investigate polariton-intrinsic transitions.

## METHODS

**Selective Dispersion of (6,5) SWCNTs and film preparation.** As reported previously,[59] (6,5) SWCNTs were selectively extracted from CoMoCAT raw material (Chasm Advanced Materials, SG65i-L58, 0.38 g L$^{-1}$) by polymer-wrapping with PFO-BPy (poly[(9,9-dioctylfluorenyl-2,7-diyl)-alt-(6,6'-(2,2'-bipyridine))], American Dye Source, M$_w$ = 40 kg mol$^{-1}$, 0.5 g L$^{-1}$) in toluene using a shear force mixing process (Silverson L2/Air, 10,230 rpm, 72 h). Impurities were removed by centrifugation at 60,000g (Beckman Coulter Avanti J26XP centrifuge) for 2 × 45 min with intermediate supernatant extraction. The resulting dispersion was passed through a PTFE membrane filter (Merck Millipore, JVWP, 0.1 µm pore size) to collect the nanotubes and remove excess polymer. The filter cakes were peeled from the PTFE membrane and washed three times with toluene at 80 °C for 15 min before 0.8 mL of a 2 g L$^{-1}$ PFO-BPy solution in toluene were added and the mixture was sonicated for 1 h. Subsequently, 0.2 mL of toluene were added in 50 µL steps, each followed by 15 min sonication until a homogeneous liquid with a honey-like viscosity was obtained and used for film formation, resulting in about 1.1 wt% of (6,5) SWCNT in the film.

**Microcavity Fabrication.** The microcavity shown in **Figure 1a** was prepared on a clean glass substrate (Schott AF32eco, 300 µm) with a 2 nm chromium adhesion layer and a 30 nm thick thermally evaporated gold bottom mirror. A spacer layer of AlO$_x$ (120 nm) was deposited by atomic layer deposition (Ultratech, Savannah S100, precursor trimethylaluminium, Strem Chemicals, Inc.) at 80 °C. The SWCNT layer (see above) was spin coated at 800 rpm followed by another AlO$_x$ (120 nm) space layer and 30 nm gold top mirror. The reference sample was prepared likewise but without gold mirrors.

**Steady State Measurements.** Transmission spectra were recorded with a V-770 (JASCO) spectrophotometer. For angle-resolved reflectivity measurements, a white light source (Ocean Optics, HL-2000-FHSA) was focused onto the sample by an infinity corrected ×100 nIR objective with 0.85 NA (Olympus, LCPLN100XIR). The resulting spot diameter of ~2 µm defined the investigated area on the sample. For angle-resolved PL, the white light source was replaced with a 640 nm laser diode (Coherent OBIS, 5 mW, continuous wave) and reflected laser light was blocked by a 850 nm cutoff long-pass filter. The reflected/emitted light from the sample was imaged onto the entrance slit of an imaging spectrometer (Princeton Instruments IsoPlane SCT 320) using a 4f Fourier imaging system ($f_1$ = 200 mm and $f_2$ = 300 mm). The resulting angle-resolved spectra were recorded with a 640 × 512 InGaAs array (Princeton Instruments, NIRvana:640ST). A linear polarizer was placed in front of the spectrometer to select between *s* and *p* polarization.

**Time-dependent PL measurements.** The spectrally filtered output of a picosecond-pulsed supercontinuum laser source (Fianium WhiteLase SC400) was focused onto the sample by an objective (Olympus, LCPLN100XIR) and imaged confocally onto an Acton SpectraPro SP2358 spectrograph (grating 150 lines mm$^{-1}$). A dichroic long-pass filter (830 nm cutoff) was used to block scattered laser light. A liquid nitrogen cooled InGaAs line camera (Princeton Instruments OMA-V) was used for spectral acquisitions required to find the desired cavity emission spectrum. The spectrally filtered PL emission was then imaged onto a gated InGaAs/InP avalanche photodiode (Micro Photon Devices) *via* a ×20 nIR optimized objective (Mitutoyo). Photon arrival time statistics were acquired by a time-correlated single-photon counting module (Picoharp 300, Picoquant GmbH). The instrument response function (IRF) was estimated by the detector-limited PL decay of a (6,5) SWCNT reference at the $E_{11}$ transition.

**Pump-probe Measurements and Global Analysis.** Femtosecond transient transmission (TT) and absorption (TA) measurements were performed with a commercial TA spectrograph (Helios Fire, Ultrafast Systems). The pump pulses were spectrally centered at 1000 nm and 576 nm and generated with a commercial optical parametric amplifier (TOPAS-Prime, Light Conversion), that was pumped by a regeneratively amplified femtosecond Ti:Sapphire laser (Astrella, Coherent) centered at 800 nm, with a 4 kHz repetition rate, 78 fs pulse durations, and 1.6 mJ pulse energy. The spot size of the focused pump beam was about 250 μm at the sample position. Typically, pump fluences were 200 μJ cm$^{-2}$ for pulse energies of 100 nJ. The supercontinuum probe beam was linearly polarized at the magic angle (54.7 °) relative to the pump polarization. Experimental spectra were corrected for the group velocity dispersion of the broad-band probe beam before analysis. All measurements were performed under ambient conditions. Time-resolved datasets were analyzed by standard global analysis as described previously.[60] The number of exponentials was chosen to minimize fitting error and residual structure.

**Data Analysis and Simulation.** The angle-resolved reflectivity data was analyzed by a coupled oscillator model as described previously.[21] In short, the UP and LP dispersions were fitted by the analytical expression for their energy eigenvalues

$$E_{UP/LP} = \frac{1}{2}(E_X - i\hbar\Gamma_X + E_C - i\hbar\Gamma_C) \pm \frac{1}{2}\sqrt{V_A^2 + 0.25(E_X - i\hbar\Gamma_X - E_C + i\hbar\Gamma_C)^2} \quad (1)$$

where $E_X$ is the SWCNT $E_{11}$ exciton energy and $E_C$ is the cavity energy dispersion, which depends on the effective refractive index $n_{eff}$ between the mirrors and the angle $\theta$ by $E_C(\theta) = E_0\left(1 - \left(sin(\theta)/n_{eff}\right)^2\right)^{-1/2}$. $\Gamma_C$ and $\Gamma_X$ are the half width at half maximum of the cavity and exciton resonance, respectively, and $V_A$ is the coupling potential, which depends on the Rabi splitting $\hbar\Omega$ with $V_A = \sqrt{\hbar\Omega^2 + (i\hbar\Gamma_C - i\hbar\Gamma_X)^2}$.

Transfer-matrix simulations of the cavity transient transmission were performed based on complex refractive index data calculated from the TA data recorded for the SWCNT reference sample using Kramers-Kronig relations (**Supporting Information**, **Figure S6**). The input parameters, *i.e.*, the layer thicknesses of the microcavity, the average interface roughness between layers, the fraction of ground state bleach and a scaling parameter, were extracted by fitting the experimental differential transmission spectrum at a pump-probe delay of 10 ps for a microcavity excited at the same pump wavelength as the reference using a genetic algorithm (**Figure S7** and **Tables S1 and S2, Supporting Information**). The quality of the fit was assessed by comparing the fitted layer thicknesses and interface roughness with experimental values obtained from atomic force micrographs (Bruker Dimension Icon, tapping mode) of reference layers fabricated under the same conditions as the cavity layers (see **Table S2, Supporting Information**).

AUTHOR INFORMATION

**ORCID**

Jana Zaumseil: 0000-0002-2048-217X

Tiago Buckup: 0000-0002-1194-0837

ACKNOWLEDGEMENTS

This research was funded by the Volkswagenstiftung (Grant No. 93404). N.F.Z. and J.Z. also acknowledge funding from the European Research Council (ERC) under the European Union's Horizon 2020 research and innovation programme (Grant agreement no. 817494, "TRIFECTs").

# Supporting Information

# Table of Contents

## 1. (6,5) SWCNT absorption and emission spectra

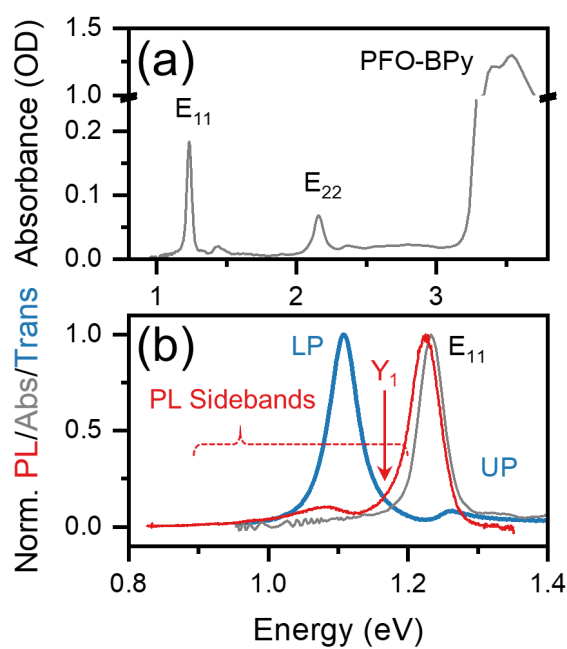

**Figure S1.** (a) Absorbance of a (6,5) SWCNT reference film in a PFO-BPy matrix. The two lowest exciton resonances ($E_{11}$ and $E_{22}$) of polymer-sorted (6,5) SWCNT are labeled. (b) Normalized emission (red) and absorbance (grey) of a (6,5) SWCNT reference film and transmission spectrum of the strongly coupled cavity (blue) with lower polariton (LP) and upper polariton (UP). The characteristic SWCNT sidebands are red-shifted to the main $E_{11}$ transition and overlap with the LP maximum. The high-angle PL contribution in Figure 1b is due to the indicated $Y_1$ sideband.

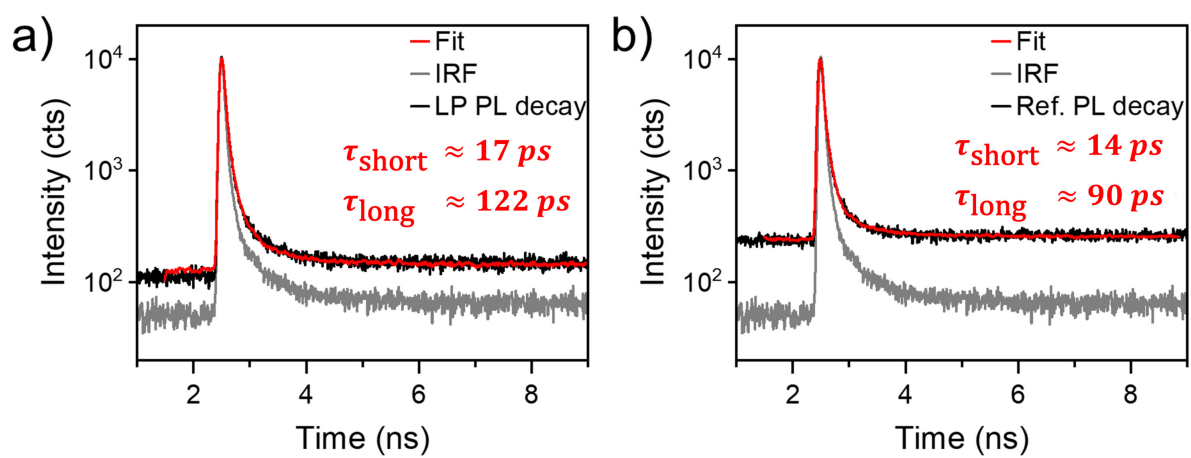

**Figure S2.** (a) LP photoluminescence decay of the sample depicted in **Figure 1**. The fit was performed with a biexponential re-convolution accounting for the IRF. (b) Photoluminescence decay of a reference (6,5) SWCNT film measured at the emission wavelength corresponding to LP energy. Both time traces show a similar decay behavior taking into account the instrument response function (IRF).

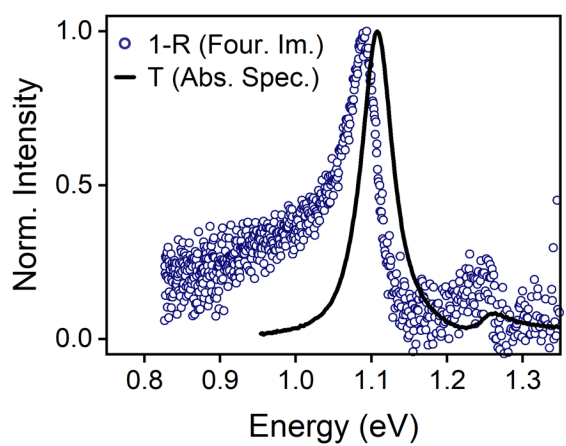

**Figure S3.** Comparison between angle-resolved reflectivity and transmission data under normal incidence. Blue open circles show the angle-resolved reflectivity data recorded by Fourier imaging (Four. Im.). The data is plotted as 1-R and normalized to be comparable to the transmission data (black line) recorded with a commercial absorption spectrometer (Abs. Spec., see methods for details). The observed deviation between reflectivity and transmission data arises from a slight thickness deviation of the spin-coated SWCNT layer and is within expectation.

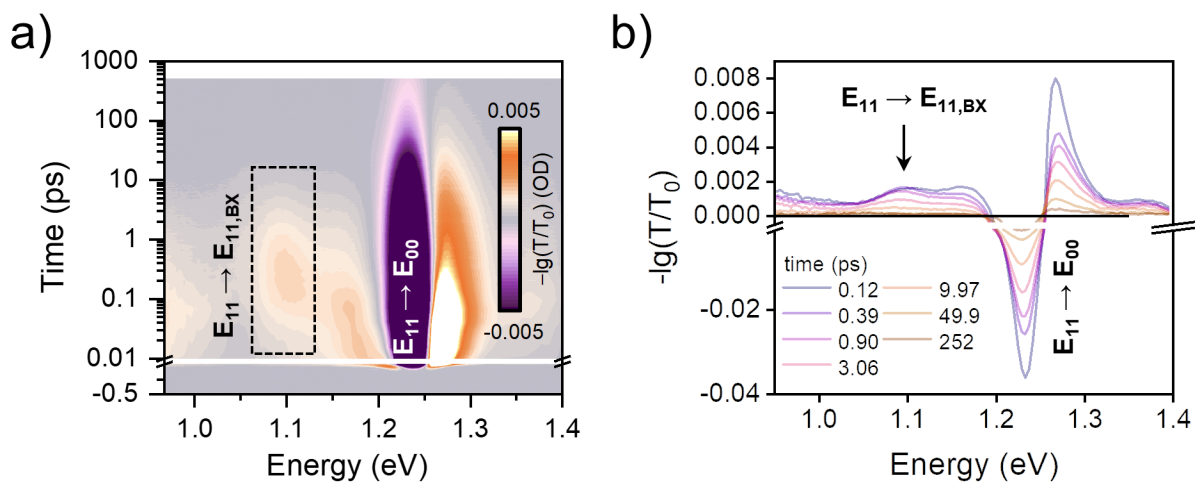

**Figure S4.** (a) Transient absorption spectra of a (6,5) SWCNT reference film in a PFO-BPy matrix. The color scale was chosen to make the biexciton transition visible. (b) Selected transient absorption spectra from (a) given in absolute differential transmission. The indicated $E_{11}$ to biexciton transition corresponds to a binding energy of 130 meV for the biexciton, as reported previously.[1] It is also consistent with transient absorptions spectra of polymer-wrapped (6,5) SWCNT dispersions.[2]

## 2. Transient complex refractive index of metal mirrors

Apart from the emitter layer, the other cavity layers, *i.e.*, oxide spacer and metal mirrors, may also exhibit transient changes in their complex refractive indices produced by the pump pulse. At the pump fluences and wavelengths employed in this work, the aluminium oxide spacer layer is not excited by the pump pulse. However, thin gold or metal layers in general, are known to be thermally excited by near-infrared pump pulses leading to the so-called thermo-reflectance response.[3] To describe this effect, we follow the adaptation of the two-temperature model for thermal excitation of metals by Liu *et al.*[4] for microcavities. The thermal excitation is described by an initial heat transfer of the absorbed pump-fluence by the gold mirror to the electron gas, which subsequently transfers the energy to the lattice (**Figure S5a**). For the duration of the increased electron-gas temperature, the damping rate of the plasma frequency is increased, which affects the dielectric response of the metal and hence the complex refractive index. This can be described by a Lorentz-Drude model for the dielectric function[4] or complex refractive index of gold (**Figure S5b**). By extracting the time-dependent change of the damping rate of the plasma frequency from the temperature transfer between electron-gas and lattice, the transient complex refractive index of the gold mirror upon excitation can be calculated (**Figure S5c**). For our experimental parameters we find the effect of thermal mirror excitation on the transient complex refractive index on differential transmission (< 1 mOD, **Figure S5d**) to be negligible compared to the overall response (**Figure 2, main text**). Furthermore, the small increase of the gold lattice temperature $\Delta T \approx 0.22\ K$ due to the pump pulse does not lead to a significant change in layer thickness $\Delta L/L$ considering the linear thermal expansion coefficient for aluminium oxide ($\alpha_L = 1.3 \cdot 10^{-5}\ K^{-1}$)[5] :

$$\frac{\Delta L}{L} = \alpha_L \cdot \Delta T = 2.9 \cdot 10^{-4}\% \tag{1}$$

Using the transfer matrix scheme, we obtain a change in transmission of about $7 \cdot 10^{-5}$ OD. Accordingly, all changes in transient transmission can be attributed to absorption by SWCNTs or polariton states in good approximation.

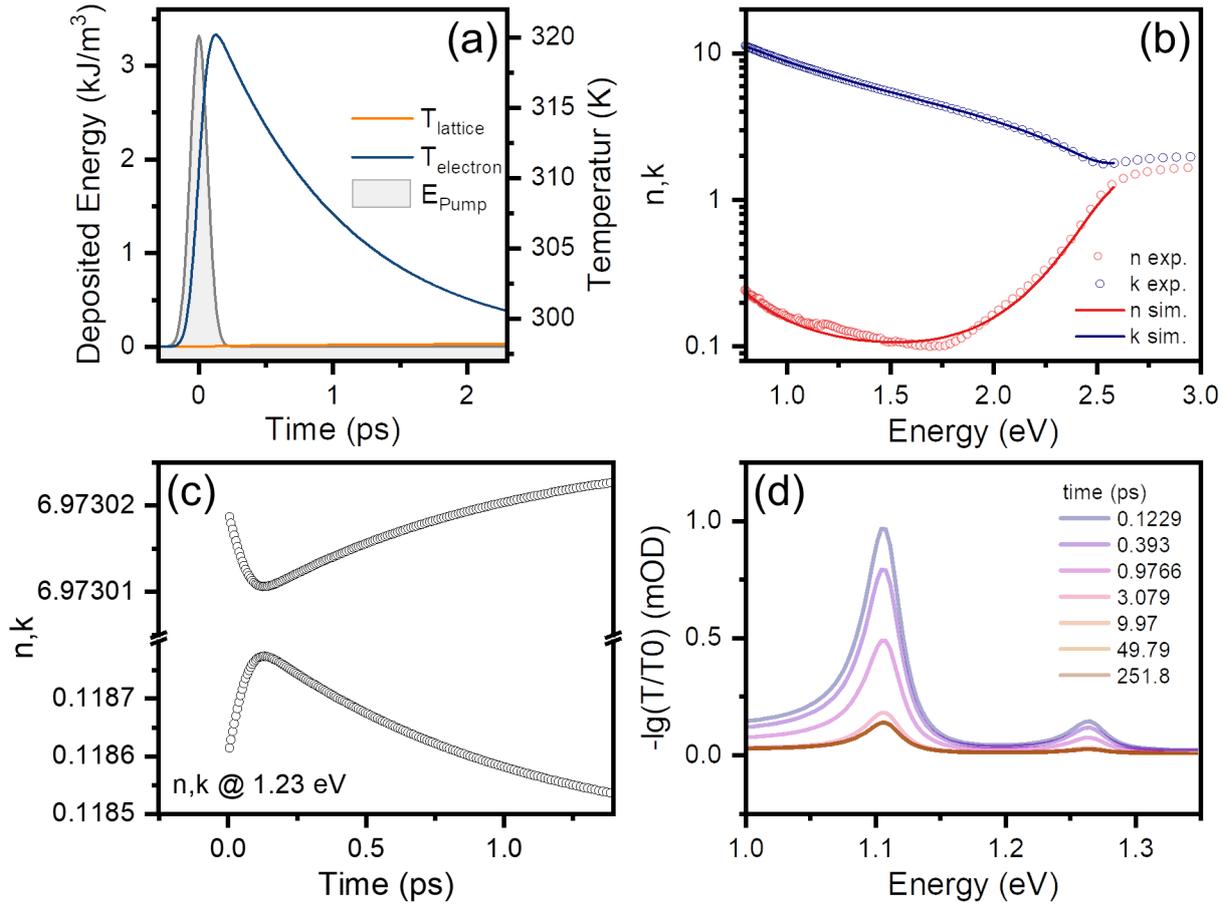

**Figure S5.** (a) Deposited energy by the pump pulse (61.1 µJ/cm$^2$, 1000 nm) compared to electron gas and lattice temperature of a 30 nm gold mirror as a function of pump-probe delay. (b) Imaginary (*k*) and real (*n*) part of the complex refractive index calculated from a multi-Lorentzian model of the complex dielectric function (solid line).[6] Open circles show the experimental complex refractive index of gold in steady state under ambient conditions.[7] (c) Transient change of the complex refractive index of a 30 nm gold mirror calculated by the two-temperature model.[4] (d) Simulation of cavity transient transmission using the transient complex refractive index for thermal excitation of the gold mirrors shown in (a)-(c) with the emitter layer excitation turned off.

## 3. Extraction of transient complex refractive index of the emitter layer

In order to calculate the transient transmission of the cavity using the transfer matrix (TM) scheme, the transient change of the complex refractive index of the emitter layer is required. To obtain this, we calculate the SWCNT absorption for different pump probe delays $t$ from the differential transient absorption (TA) spectrum of the SWCNT reference film, given in $-log_{10}(A(t)/A_0)$, by adding the steady state absorption $log_{10}(A_0)$:

$$log_{10}(A(t)) = log_{10}(A_0) + log_{10}(A(t)/A_0) \qquad (2)$$

and convert it into the imaginary part of the complex refractive index $k$ (see **Figure S6a**)

$$k(t) = \frac{log_{10}(A(t)) \cdot \lambda}{log_{10}(e) \cdot 4\pi \cdot d(SWCNT)} \qquad (3)$$

where $\lambda$ is the wavelength, $e$ is Euler's number and $d(SWCNT)$ is the average thickness of the SWCNT reference film as obtained by profilometry. Using the Kramers-Kronig relation for the complex refractive index, we can calculate the real part $n$ of the refractive index (**Figure S6b**) as described previously.[8] The results correspond well to the complex refractive index of (6,5) SWCNTs embedded in a PFO-BPy matrix as measured with ellipsometry (**Figure S6b**, brown dashed line) by Graf et al.[8]

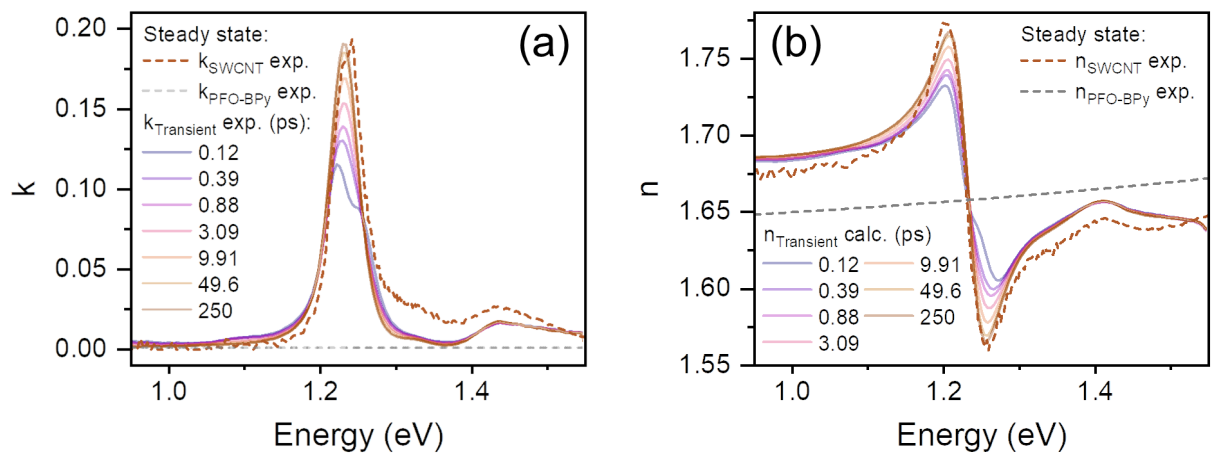

**Figure S6.** (a) Imaginary part ($k$) of the complex refractive index calculated from the TA data of the SWCNT reference at different delay times (solid lines). (b) Real part of the complex refractive index ($n$) calculated with the Kramers-Kronig relation at different time delays (solid lines). Corresponding imaginary and real part of the complex refractive index for a (6,5) SWCNT/PFO-BPy film (brown dashed line) and a PFO-BPy film (grey dashed line) measured in steady state by ellipsometry are given for comparison.

# 4. Genetic algorithm-assisted transfer matrix simulation of transient transmission data

The transient change in cavity transmission is simulated by the transfer matrix (TM) scheme adapted from Pettersson et al.[9] The transfer matrix scheme predicts the amplitudes of transmitted and reflected components for an electromagnetic plane wave that is incident on an arbitrary material stack by solving Fresnel's equations. In first approximation, that means assuming homogeneous layers without interface roughness. The required inputs are the layer thicknesses and material complex refractive indices in steady state. To match the observed experimental linewidths, we also account for interface roughness by including scalar scattering according to Yin et al.[10] For simplicity, we use a global scattering parameter σ describing the root mean square (RMS) roughness averaged over all interfaces. To include the time dependence, we use the transient complex refractive index of the emitter layer obtained by the TA spectrum of a SWCNT reference layer as described in the previous section. Note that for our experimental conditions, only the refractive index of the (6,5) SWCNTs changes significantly as a function of time. To account for the reduced pump fluence due to the cavity top mirror (~ 95% loss at 1000 nm), we use SWCNT TA data recorded for a 2 nJ pump pulse to simulate a cavity excited with a 30 nJ pump pulse, (~ 93% loss at 1000 nm).

The major issue of this approach is firstly, that the morphology and size of the sample area investigated by pump-probe spectroscopy may slightly differ from the morphology and size of the sample area measured in steady state transmission, which may lead to deviations when calculating the transient refractive index with equation (3). Secondly, the emitter layer thickness and film roughness can only be estimated by profilometry measurements, which may not precisely resemble the morphology of the sample area in pump-probe spectroscopy. This reduces the agreement between simulated and experimental cavity TA data, as can be seen in previous reports.[11, 12] We remedied these issues by combining the TM scheme with a commercial genetic algorithm (GA)[13] to optimize the set of input parameters for a given experimental spectrum.

The GA randomly tests different values for the input parameters, and compares the resulting TA spectrum to the reference spectrum to optimize the input parameters. **Figure S7a** shows the fitting routine scheme. The possible parameter values have to be given as vectors,[13] with the smallest entry being the lower bound and the largest entry being the upper bound of the parameter. In contrast to a conventional regression method, the best fit value is limited by the coarseness of the parameter values. Here, we used standard settings for the GA (given in **Table S1**). The population size is the number of parameter combinations tested in each iteration, referred to as generations. The crossover fraction determines the number of parameter combinations that were selected from previously successful combinations. The number of stall generations gives the number of iterations allowed, that do not lead to further improvement of the best parameter combination. The settings may be adapted according to problem size and computation power. Note that the GA randomly tests the parameter space, which is why the final result may slightly differ for each run.

**Table S1.** Genetic Algorithm settings.

| Parameter | Population size | Generations | Crossover fraction | Stall generations |
|---|---|---|---|---|
| Value | 200 | 100 | 0.5 | 40 |

To test this approach, we calculated the TA spectrum of the reference (6,5) SWCNT film using the GA assisted TM simulation at 0.5 ps pump-probe delay (**Figure S7b**). The obtained SWCNT film thickness of 40.3 nm lies within the uncertainty estimated by profilometry (30 $\pm$ 15) nm and the obtained roughness of 27 nm is close to the experimentally determined local RMS roughness of 10 nm for the reference film. Note, that without these degrees of freedom, the residual between simulated and experimental data is indeed higher, due to the uncertainties of the input parameters.

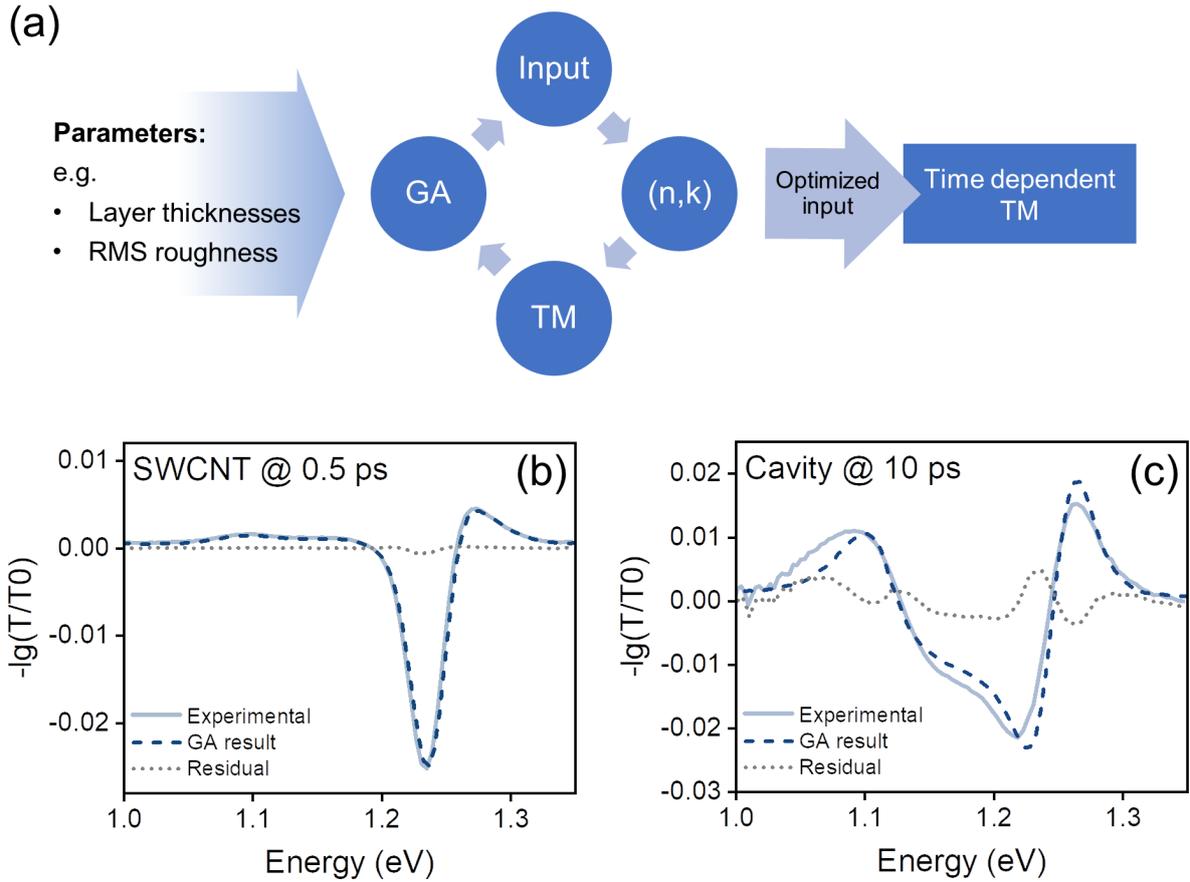

**Figure S7.** (a) Schematic illustration of the GA assisted TM simulation of transient transmission data. (b) GA assisted TM simulation of the SWCNT TA spectrum for the reference film at 0.5 ps pump-probe delay. (c) GA assisted TM simulation of the cavity TA spectrum under UP excitation for 10 ps time delay. The GA results for the input parameters are given in **Table S2**.

For the cavity transient transmission (TT) data, a similar problem arises, that is, the precise film thickness of the cavity mirrors and spacer layers at the investigated sample area can only be estimated from local profilometry measurements of reference layers or samples. Using the GA assisted TM simulation, we can optimize the input parameters to fit the experimental data. **Figure S7c** shows simulated and experimental cavity TT data for UP excitation. As a reference spectrum for the cavity TT data we chose 10 ps pump-probe delay, for which spectra should be governed by the recovery of the SWCNT bleach while at the same time the experimental spectrum exhibits a reasonable signal to noise ratio in regions of low transmission. For parameter bounds well beyond the experimental uncertainty of the respective parameter, we find that the algorithm converges at values close to the experimental values (**Table S2**). A detailed description of the parameters shown in **Table S2** is given below.

**Table S2**. Summary of GA results for the simulation of the TT cavity data for UP excitation at 10 ps pump-probe delay.

| Parameter | SWCNT thickness (nm) | Spacer thickness (nm) | Mirror thickness (nm) | RMS roughness (nm) | Bleach scaling parameter | Global scaling parameter |
|---|---|---|---|---|---|---|
| Lower bound | 15 | 145 | 20 | 1 | 0.01 | 0.01 |
| Upper bound | 55 | 95 | 35 | 200 | 100 | 100 |
| Interval | 0.5 | 0.5 | 0.5 | 1 | 0.01 | 0.01 |
| GA Result | 46 | 116 | 28.5 | 22 | 1.53 | 0.95 |
| Experimental | 30 ± 15 | 125 ± 2 | 30 ± 1* | 10 | 1.41† | |

*From internal evaporator calibration

†From $E_{11}$ to $E_{22}$ absorption ratio.

The uncertainty of the SWCNT layer thickness corresponds to the variation found for a reference layer and was calculated from three height measurements at sample positions spaced over a distance of 1 cm. The RMS roughness is that of a (6,5) SWCNT layer determined from a 5 x 5 μm atomic force micrograph. All profilometry data was obtained from atomic force micrographs (AFM, Bruker Dimension Icon, tapping mode). Values marked with an asterisk result from local profilometry measurements, which may explain the higher deviation from the GA results, as they do not represent the whole sample. The mirror layer thickness was estimated from the internal calibration of the thermal evaporator with a uniformity of ±3%. The bleach scaling parameter was included to account for a more efficient excitation close to the $E_{11}$ wavelength and equals approximately the ratio between the $E_{11}$ and $E_{22}$ peaks in absorption (**Table S2**, dagger). For the simulated cavity data, we included a global scaling parameter, to match the experimental data more closely for better comparison. Without this parameter, we obtain slightly higher values for the simulated differential transmission peaks (+ 5-10%) compared to the experimental spectra.

## 5. Transient transmission data

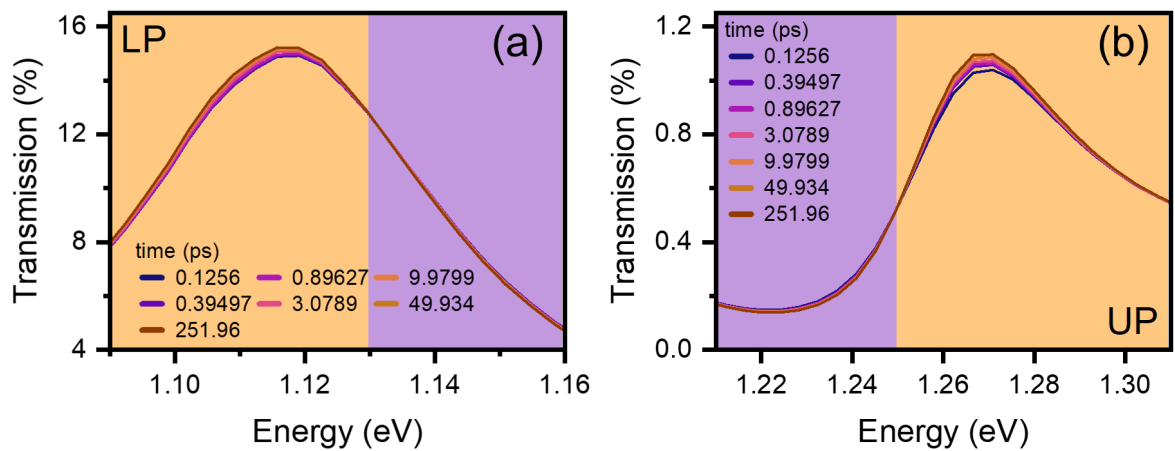

**Figure S8.** TM simulation of the polariton mode shift in transmission for different pump-probe delays for (a) the LP mode and (b) the UP mode. Increased and decreased transmission is indicated in orange and purple, respectively. The simulation parameters were obtained by a genetic algorithm fit to the experimental differential transmission data at 10 ps. Note that the data is given in transmission (not differential transmission) for better illustration.

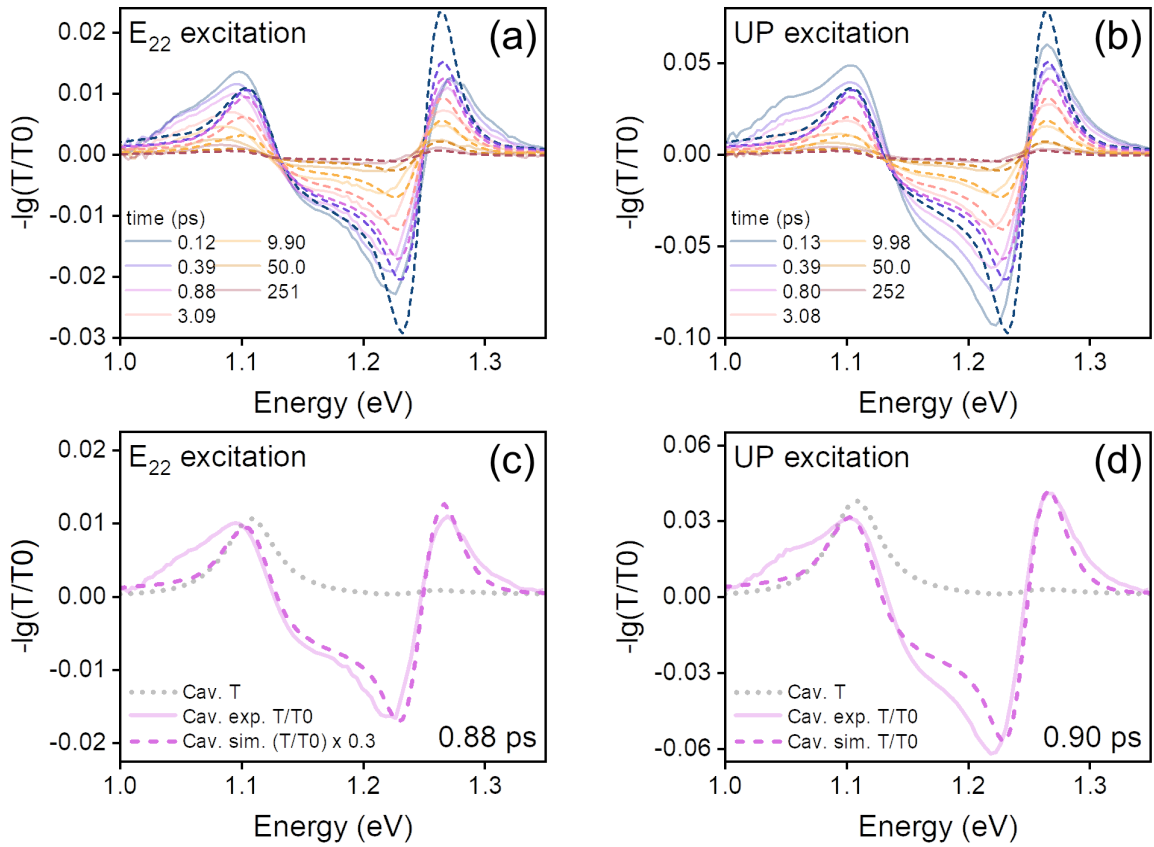

**Figure S9.** (a) Experimental (solid lines) and simulated (dashed lines) transient transmission of a strongly coupled microcavity under non-resonant $E_{22}$ excitation. (b) Experimental (solid lines) and simulated (dashed lines) transient transmission of a strongly coupled microcavity under resonant UP excitation. The simulation parameters were obtained by a genetic algorithm fit to the experimental data at 10 ps. The transient spectra around 0.9 ps together with the steady state cavity transmission are shown (c) and (d).

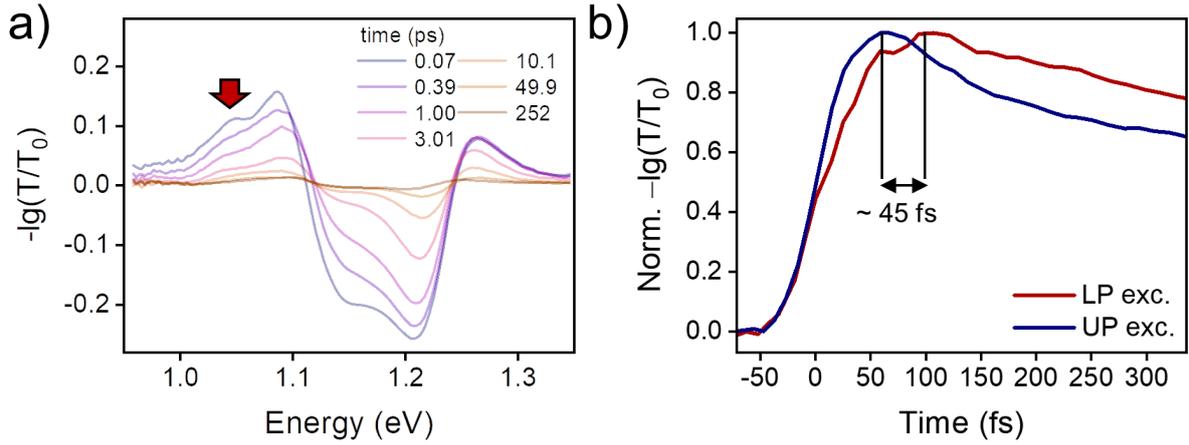

**Figure S10.** (a) Cavity TT spectrum for LP excitation (1.09 eV). The red arrow marks the shoulder of the proposed UP to biexciton transition. Note that the overall intensity is increased by about a factor of three owing to the large intensity of the LP compared to the UP. (b) Time traces at the UP to biexciton transition (1.05 eV) for LP (red) and UP (blue) excitation. The delay in maximum transition intensity is 45 fs. The origin of this long delay is unclear.

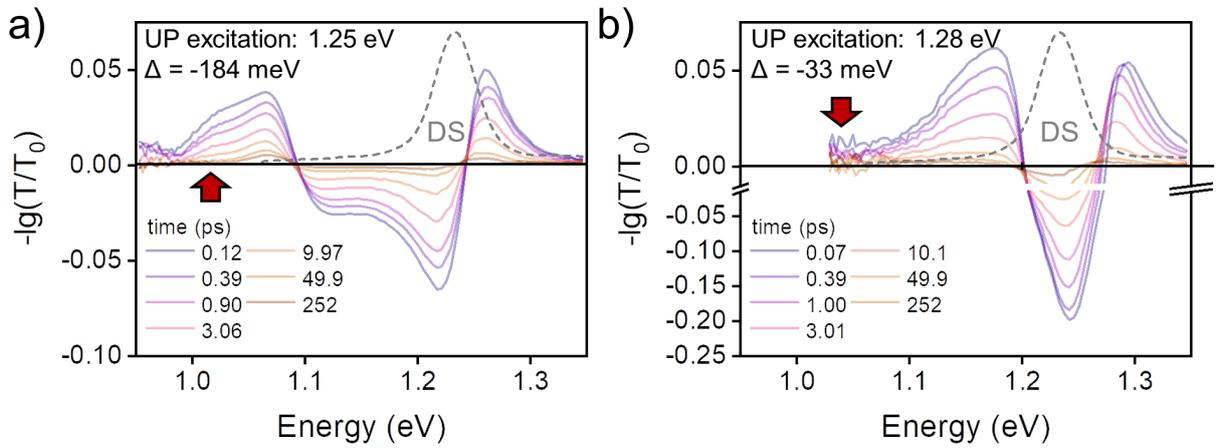

**Figure S11.** UP excitation for cavities for two different detunings. (a) UP excitation for a cavity with increased detuning (-184 meV) compared to the data in Figure 2 (-140 meV). The red arrow indicates where the UP to biexciton transition is observed. (b) UP excitation for a cavity with decreased detuning (-33 meV) compared to the data in Figure 2. The red arrow indicates where the UP to biexciton transition would be expected. The dashed grey line indicates the (6,5) SWCNT film absorption, which can be taken as a measure for the dark states (DS) or exciton reservoir density of states. Note that the data in (b) was only measured until 1.04 eV owing through the limited transmission at lower energies for this sample. The appearance of the transient spectra of (b) is changed compared to (a) as the negative parts of the LP and UP ground state responses overlap at this detuning.

## 6. Kinetic model

The kinetic model proposed for Figure 4 of the main text can be used to estimate the evolution of the UP population under different pumping conditions. The kinetic equations are:

$$N_{UP}(t)/dt = k_{DSUP}N_{DS}(t) - k_{UPDS}N_{UP}(t) \quad (4)$$

$$N_{DS}(t)/dt = k_{UPDS}N_{UP}(t) + k_{E_{22}}N_{E_{22}}(t) - (k_{LPDS} + k_{DSUP} + k_{nr})N_{DS}(t) \quad (5)$$

$$N_{LP}(t)/dt = k_{DSLP}N_{DS}(t) - k_{Ph}N_{LP}(t) \quad (6)$$

$$N_{E_{22}}(t)/dt = -k_{E_{22}}N_{E_{22}}(t) \quad (7)$$

where $N_{UP}$, $N_{DS}$, $N_{LP}$, and $N_{E22}$ are the populations in the UP, dark states (DS), LP and $E_{22}$ states respectively. $k_{DSUP}$ describes the relaxation from the UP to DS, this process should be very fast and for simplicity we take it to be $(15\ fs)^{-1}$. $k_{UPDS}$ describes the population from DS to UP, this process should be significantly slower, and we take it to be $(150\ fs)^{-1}$, as reported by Virgili et al.[14] Here, $k_{LPDS}$ is the relaxation from dark states to LP. The LP is radiatively pumped, therefore $k_{LPDS}$ is limited by the radiative decay of the SWCNTs, which is $(1\ ns)^{-1}$.[15] Because the radiative decay from the LP follows the dynamics of the SWCNT reference (compare **Figure S2**), the dark states must share the non-radiative decay channel of weakly coupled SWCNTs, therefore, we take the non-radiative deactivation of the dark states as $k_{nr} = (1\ ps)^{-1}$. The LP decays radiatively with approximately the photonic decay rate, $k_{Ph} = (15\ fs)^{-1}$. In case of $E_{22}$ excitation, the DS are populated from SWCNT. We assume in this model, that the rate of $E_{22}$ to $E_{11}$ relaxation is similar to the $E_{22}$ to DS relaxation. For the $E_{22}$ to $E_{11}$ we take a rate of $k_{E_{22}} = (100\ fs)^{-1}$, corresponding to the average rate found for different SWCNT systems.[2]

Excitation can be described by a Gaussian generation term (modeling the uncertainty in pumping and probing of the population) with a pulse width $w$ of 90 fs, by adding a term $f(t)$ of the form:

$$f(t) = (2\sqrt{\log 2}/\sqrt{\pi} \cdot w) \cdot \exp(-4 \log 2\ (t/w)^2) \quad (8)$$

The generation term is added to the state, which is excited, that is UP (eq. 4) or $E_{22}$ (eq. 7). We can numerically solve the equations 4-7 together with eq. 8 for the excitation, with finite time intervals of 2 fs. The results are shown in **Figure S12**.

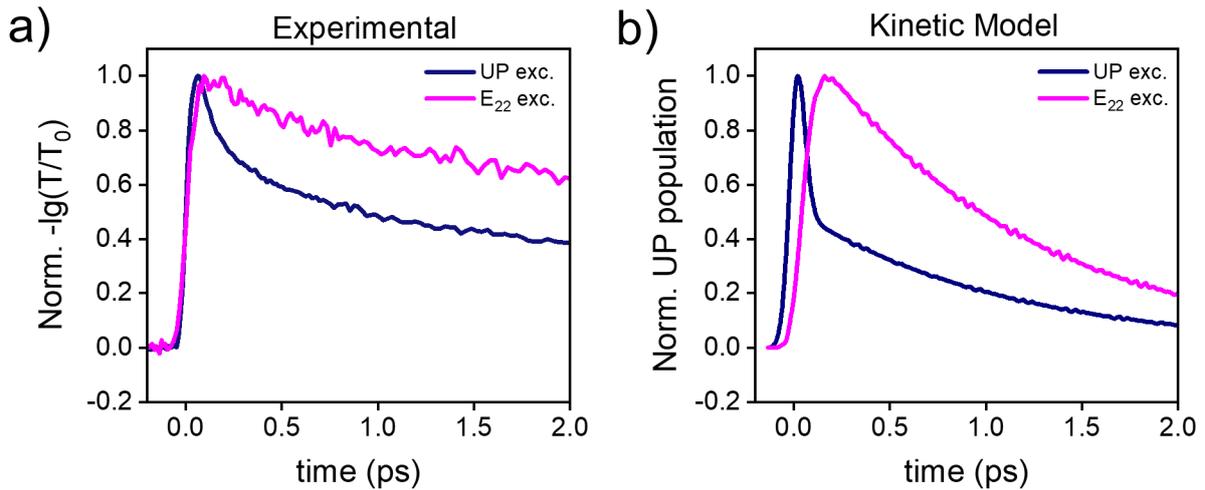

**Figure S12.** a) Experimental time traces of the proposed UP to BX transition for UP and $E_{22}$ excitation as shown in Figure 4 of the main text. b) Normalized UP population extracted from a numerical solution of the differential equations (eq. 4-7) together with a generation term (eq. 8) for the pumped state.

## 7. Biexciton transition efficiency

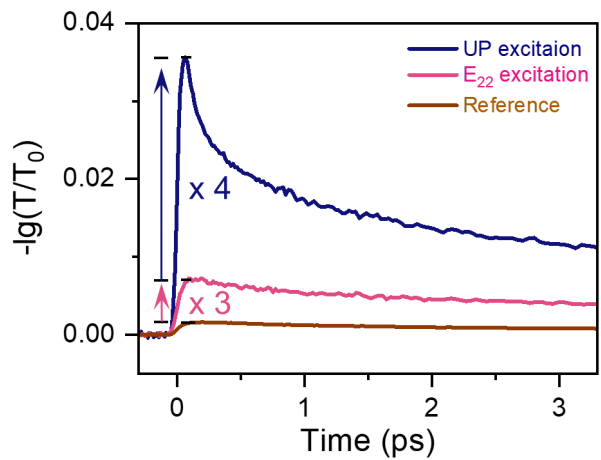

**Figure S13.** Time traces of the proposed UP to biexciton feature for a cavity (pink, blue) and a reference sample (brown) excited at the $E_{22}$ transition under equivalent excitation fluences. The photoinduced absorption feature of the cavity sample is three times more intense than for the reference sample in case of $E_{22}$ excitation. UP excitation is four times more efficient than $E_{22}$ excitation owing to the transient population of the UP.

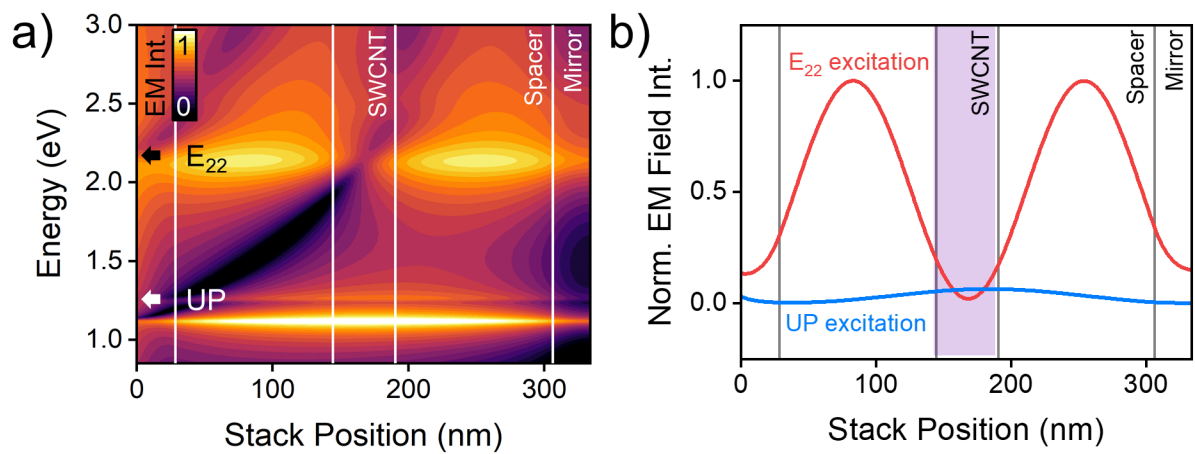

**Figure S14.** (a) Simulated electromagnetic field intensity inside the strongly-coupled cavity as a function of excitation energy and stack position. The layer thicknesses were extracted using the GA-assisted TM simulation (see above). (b) Overlap between the electromagnetic field and the SWCNT layer for UP and $E_{22}$ excitation energies. From the integrated electric field inside the SWCNT film, we find that $E_{22}$ excitation is about 17% more efficient than UP excitation at this detuning.